\documentclass[twocolumn,preprintnumbers,superscriptaddress,amsmath,amssymb,pre]{revtex4}

\usepackage{graphicx}
\usepackage{dcolumn}
\usepackage{subfigure}
\usepackage{bm}
\usepackage[utf8]{inputenc}  
\usepackage{comment}

\usepackage[cmyk,dvipsnames]{xcolor}
\definecolor{pacificb}{HTML}{1CA9C9}

\def \be {\begin{equation}}
\def \ee {\end{equation}}
\def \beA {\begin{eqnarray}}
\def \eeA {\end{eqnarray}}

\def \Im  {\rm{Im}}

\begin{document}
\title{Modelling reservoir computing with the discrete nonlinear Schr\"odinger equation}

\date{\today}

\author{Simone Borlenghi} 
\affiliation{Department of Applied Physics, School of Engineering Sciences, KTH Royal Institute of Technology, Electrum 229, SE-16440 Kista, Sweden}
\author{Magnus Boman}
\affiliation{KTH Royal Institute of Technology, EECS/SCS, Electrum 229, SE-16440 Kista, Sweden}
\affiliation{RISE SICS, Electrum 230, SE-16429 Kista, Sweden}
\author{Anna Delin}
\affiliation{Department of Applied Physics, School of Engineering Sciences, KTH Royal Institute of Technology, Electrum 229, SE-16440 Kista, Sweden}
\affiliation{Swedish e-Science Research Center (SeRC), KTH Royal Institute of Technology, SE-10044 Stockholm, Sweden}

\begin{abstract}

We formulate, using the discrete nonlinear Schr\"odinger equation (DNLS), a general approach to encode and process information based on reservoir computing. Reservoir computing is a promising avenue for realizing neuromorphic computing devices.  In such computing systems, training is performed only at the output level, by adjusting the output from the reservoir with respect to a target signal. In our formulation, the reservoir can be an arbitrary physical system, driven out of thermal equilibrium by an external driving. The DNLS is a general oscillator model with broad application in physics and we argue that our approach is completely general and does not depend on the physical realisation of the reservoir. The driving, which encodes the object to be recognised, acts as a thermodynamical force, one for each node in the reservoir. Currents associated to these thermodynamical forces in turn encode the output signal from the reservoir. As an example, we consider numerically the problem of supervised learning for pattern recognition, using as reservoir a network of nonlinear oscillators.

\end{abstract}

\maketitle

Present computers consist of two main separate parts: the central processing unit (CPU) and a memory. However, this basic design is limiting the capacity of today’s computers since CPU speed and memory size have improved much more than the communication speed between them. 

This problem is known as the von Neumann bottleneck problem \cite{backus77}. To solve it, alternative computer architectures, where co-localized memory and computation is a central part of the design, are currently being intensly explored. Both biological and artificial neural processing systems, i.e. so-called neuromorphic computing systems, are characterized by such co-localization \cite{indiveri15}, and therefore neuromorphic computing is an interesting route to explore.  In addition to providing a potential solution of the bottleneck problem, neuromorphic computing offers other advantages, such as the ability to be trained and better energy effciency \cite{hoppensteadt17}

Nonlinear oscillators provide interesting possible physical implementations for analog neuromorphic computers and a multitude of ways to do this have been suggested.
Examples include spin systems\cite{nikonov15,torrejon17} optoelectronics devices, and electric circuits \cite{hoppensteadt99,fan15,maffezzoni15,pourahmad17}.

\begin{figure}
\begin{center}
\includegraphics[width=7.0cm]{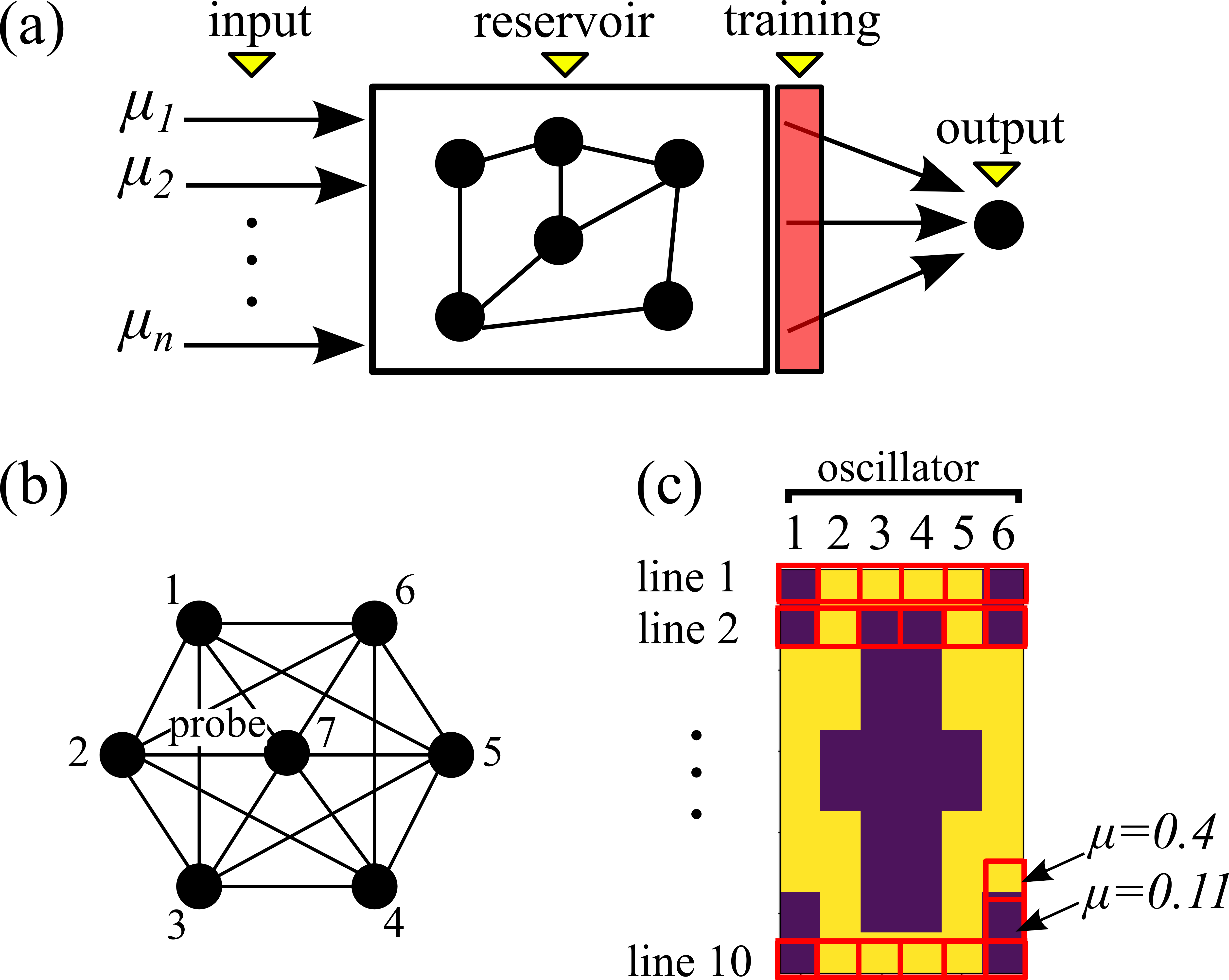}
\caption{a) Schematic of a reservoir computer, consisting of an ensemble of artificial neurons (in our case, coupled oscillators) 
driven by the external inputs $\mu_1,...\mu_N$. Those inputs encode the image to be represented.
The reservoir generates a response signal, which is sent to a passive node for the readout. The training take place only at the readout level.
b) Our reservoir is a DNLS network of 7 coupled oscillators with all-to-all coupling. The input consists of chemical potentials that excite the dynamics of oscillators 1-6. 
The output signal is the total particle current flowing from the network to oscillator 7, which acts as passive node. c) Representation of the digit 0, consisting of 
60 pixels. The chemical potentials $\mu=0.4$ and $\mu=0.11$ represent respectively yellow and dark red pixels, and the image is constructed one line at a time. }
\label{fig:figure1}
\end{center}
\end{figure}
Reservoir computing is an efficient way to implement neuromorphic computing \cite{lukosevicius09}. A reservoir computer consists of a network of artificial neurons with arbitrary connections, that form a nonlinear dynamical system called the reservoir. Some input is passed to the reservoir, which creates a higher dimensional representation of the input and generates an output signal. The training for supervised learning is performed only at the output level, while the connections between the neurons of the reservoir is decided at the beginning and never changes.
 A schematic of a reservoir computer is shown in Fig.\ref{fig:figure1}a). An interesting aspect of this approach is that the reservoir can be essentially a black box: one needs only to know the output corresponding to the input forces, and not the detailed dynamics of the network. This provides a great simplification with respect to traditional neural network setups, since changing the connections in an artificial system in real time is particularly difficult.  
Several types of reservoir computers have been studied using a variety of physical systems \cite{schrauen07}.

In this paper, we move a step forward in the field of reservoir computing by describing a radically new and very general method to encode and process information. The main idea is that the reservoir is driven out of thermal equilibrium by an external driving, which encodes the image to be recognised and acts as a thermodynamical force. The signal generated by the reservoir is encoded in the thermodynamical currents associated to those forces. The training is performed only at the output level, by adjusting the current
with respect to a target signal.  Note that by out-of-equilibrium setup we do not mean necessarily a system with a thermal gradient. Instead, the procedure is completely general and can be applied to virtually \emph{any dissipative system} controlled by an arbitrary driving. The driving can be the gradient of some potential function (such as a voltage or temperature difference), or an arbitrary non-potential force, and the associated currents refer to any transport phenomena generated by such force (such as an electrical, spin or heat current). 

There are several advantages of processing information using thermodynamical forces and currents. At first, close to thermal equilibrium, the currents are in general proportional to the forces and are therefore easy to control and detect.  Furthermore, the same concepts of thermodynamical forces and currents can describe transport in classical and quantum systems that can be very different in nature, length and time scales.
Particularly interesting are systems with several thermodynamical currents, such as spin or thermoelectric devices. In those cases the currents can be controlled independently \cite{borlenghi16}, providing very flexible and energy efficient setups.

To demonstrate the power of our approach, we here consider a reservoir consisting of a network of seven complex-valued oscillator equations, for the supervised task of image recognition. The dynamics of the network is described by an equation that generalises the discrete nonlinear Schr\"odinger equation (DNLS) \cite{iubini12}. 
A multitude of physical systems can be described by the DNLS.
Examples include micromagnetics systems, Bose-Einstein condensates (BEC), mechanical oscillators, photonics waveguides and electric circuits. 

The reservoir, depicted in Fig.{\ref{fig:figure1}b)}, is described by the following DNLS-like model

\beA\label{eq:dnls}
\dot{\psi}_m &=&i\omega_m\psi_m+[\mu_m(t)-\Gamma_m(p_m)]\psi_m\nonumber\\
                    &+& Am\sum_n\psi_n.
\eeA

for the complex amplitude $\psi_m=\sqrt{p_m(t)}e^{i\phi_m(t)}$. The first term on the right hand side is the local frequency, while the second and third terms are respectively the chemical potential and damping. The damping depends on
the local powers $p_m=|\psi_m|^2$ and explicitly reads $\Gamma_m(p_m)=\Gamma_{0m}(1+2p_m)$. For brevity we will omit the explicit dependence on $p_n$.

The chemical potential $\mu_m$, which is nonzero only for oscillators $m=1,...,6$, is the driving term that acts as a torque compensating the damping. 

The seventh oscillator is a passive node,
or probe, where the output is collected. The nonlinearity of the damping guarantees that the system has limit cycle oscillations whenever $\mu_m>\Gamma_{0m}$.
In spin systems, the chemical potential corresponds to spin transfer torque, which leads to steady state precession of the magnetisation. We shall see that, for pattern recognition, it is important to consider a time dependent chemical potential 
which drives the system into a chaotic regime. In spin systems, this can be realised using an ac current that gives a time dependent spin torque \cite{naletov11}. 

The coefficient $A_m=h_m(1-i\Gamma_{0m})$ is the dissipative coupling \cite{iubini13} that depends on each oscillator..

In the presence of different chemical potentials at the sites, the system reaches a non-equilibrium steady state where the "particle" current  $j^p_{mn}$ flows between the oscillators $m$ and $n$.

To obtain the explicit expression for the particle current, one proceeds as in the case of the probability current for the Schr\"odinger equation \cite{iubini13,borlenghi15a}. 
This gives the continuity equation for the spin power $p_m$ 
\be\label{eq:continuity}
\dot{p}_m=(\mu_m-\Gamma)_mp_m+\sum_nj^p_{mn}.
\ee
The first term on the right hand side acts as a sink or source of excitations, depending if the sign of $\mu_m-\Gamma_m$ is respectively negative or positive.
The second term corresponds to the particle current, defined as 

\be\label{eq:current}
j_{mn}^p=2{\Im}[A_m\psi_m\psi_n^*]. 
\ee
The current in the phase-amplitude representation reads $j^{p}_{mn}=2{\Im}[A\psi_m\psi_n^*]=2h\sqrt{p_mp_n}\sin[\phi_m(t)-\phi_m(t)-\beta_{mn}]$.
The term $\beta_{mn}=\arg[(1-i\Gamma_{m})h_m]$, which stems from the condition of dissipative coupling, is a $U(1)$ lattice gauge field \cite{borlenghi16a} determined by the topology of the network. When the oscillators are phase-synchronised, $\phi_m(t)=\phi_n(t)$, the current is proportional to $\sin\beta_{mn}$. The dissipative coupling guarantees that the current is not zero in the synchronised regime \cite{borlenghi15a}. Note that one does not need complete synchronisation to obtain a non-zero current. 
In particular, in noisy systems the synchronisation is broken from time to time due for example to thermal fluctuations. In this case the relevant observable could be the time-averaged current.  

The output signal used to recognise images is the total particle current reaching the probe node (oscillator 7 in Fig.\ref{fig:figure1}b), i.e. $\mathcal{J}=\sum_{n=1}^6j^p_{n7}$.

We remark that the particle current can describe several transport phenomena, such as the flow of bosons in BEC \cite{eilbeck03} or the dynamics of excitons \cite{iubini15}. In spin systems, it corresponds to the spin-wave current that describes the transport of the magnetisation component along the quantisation axis \cite{borlenghi15b}.

Hereafter, we apply the DNLS model as given in Eq.(\ref{eq:dnls}) to represents several digits, as displayed in Fig.\ref{fig:figure1}.
For each line of the image, the input is passed as a series of chemical potentials $(\mu_1,...,\mu_6)$, representing the image to be recognised. 
The value $\mu_m=0$ corresponds to a yellow pixel, while the value $\mu_m=3$ to a dark-red pixel. A schematic of the procedure is shown in Fig.\ref{fig:figure1}c). 
For each line, the six oscillators evolve in a non-equilibrium state, and the response spin current $\mathcal{J}$ as a function of time is measured at node 7. 
The image is constructed by processing one line at a time, and the total response is the sum of the current over each line. In a completely parallel and faster setup, one should use one oscillator for each pixel. 
The other extreme would be to use only one node with fading memory, that can work as a completely serial computer \cite{appeltant11}. Our method is in between parallel and serial information processing, which aims at having an optimal trade-off between speed and simplicity. 

\begin{figure}
\begin{center}
\includegraphics[width=7.0cm]{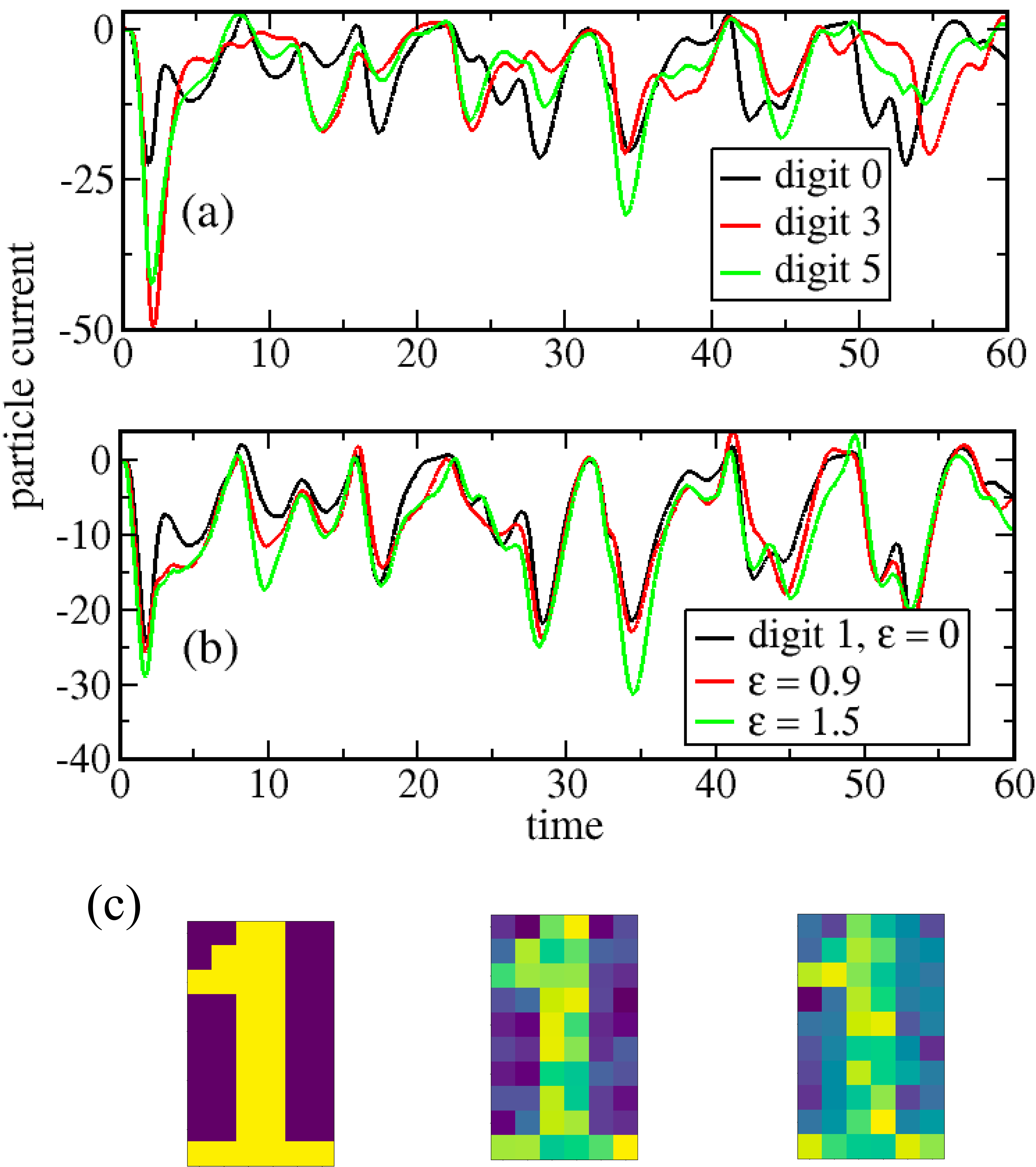}
\caption{Representations of the digits 0,3 and 5 using a binary pattern with 60 bits. Panel a) shows the time signal corresponding to those digits. 
while panel b) and c) shows the time signal for the digit 1  in the presence of noise, with intensity $\varepsilon=0,0.9$ and 1.5. The current is expressed in unit of coupling $h$.}
\label{fig:figure2}
\end{center}
\end{figure}

The simulations were performed by solving numerically Eq.(\ref{eq:dnls}) using a 4th order Runge-Kutta algorithm with integration time step $dt=0.001$ model units and a total of 6000 steps for each line of the image.
The coupling $h_1,...,h_7$ between oscillator $m=1,..,7$ and all the other oscillators has the values $(0.1,0.4,0.8,0.05,1,1.1,0.35)$, while the damping rate is $\Gamma_{0m}=0.3$. The oscillators have different proper frequencies $\omega_{01},...,\omega_{07}=(1,1.4,2,0.5,1.6,0.3,2.1)$.
The chemical potentials for the $m$th oscillator, $\mu_m(t)\equiv \mu_mC_m\sin\omega_{dm}t$ is multiplied by the time dependent factor $C_m\sin\omega_{\rm{d}m}t$, with amplitudes $C_{1},...,A_{7}=(5,15,10,12,8,4.7,2)$ and driving frequencies
$\omega_{d1},...,\omega_{d7}=(1,1.2,1.5,0.5,0.3,2,0.8)$  
Note that with this time dependent chemical potential  the system is chaotic. In this way a small difference in the input chemical potential gives different well defined outputs. This guarantees that the system has the necessary separation and approximation 
properties to recognise the digits.

Next, we test the performance of our system by studying the recognition of a noisy image. To this end, we consider a modified chemical potential $\mu_m\rightarrow \mu_m+\varepsilon\theta_m$, where $\theta_m$ is a 
Gaussian random variable with zero average and unit variance and the parameter $\varepsilon$ controls the strength of the noise. 
The signals corresponding to various digits in the absence of noise are displayed in Fig.\ref{fig:figure2}a), while the degradation of the digit 1 as a function of the noise is shown in Fig.\ref{fig:figure2} b) and c).

At this point, one needs a method to train the system to recognise images in the presence of noise. In practice, one generates an ensemble of noisy images that are compared with the target noiseless image.
As customary, we consider a system that evolves in discrete time steps $t_k$, $k=1,...,K$.
At each time step $t_k$, let the target and noisy currents be respectively $\mathcal{J}_q^T(t_k)$ and $\mathcal{J}_q^N(t_k)$. Here the subscript $q=0,1,2$ refers to the digit represented. 
At the output level, one modifies the noisy current as $\mathcal{J}_q^N(t_k)\rightarrow A_P+B_P\mathcal{J}_q^N(t_k)$, by choosing the coefficients $A_P$ and $B_P$ in such a way that it is as close as possible
to the target current. In spin systems, this can be achieved by first converting the spin current into an electrical current through the ISHE, and then modifying this current through an operational amplifier.
Mathematically, this is equivalent to finding the coefficients $A_P$ and $B_P$ that minimise the distance $\mathcal{L}(\mathcal{J}_q^T,\mathcal{J}_q^N)$ between the target and noisy current. Such a distance, or loss function, is defined by 
\be\label{eq:loss}
\mathcal{L}(\mathcal{J}_q^T,\mathcal{J}_q^N)=\sum_{k=1}^K[\mathcal{J}_q^T(t_k)-A_P-B_P\mathcal{J}_q^N(t_k)]^2,
\ee
which can be done through simple linear regression. Such regression has to be performed for each of the $p=1,...,P$ noisy images. The coefficients of the regression are given by the averages 
$A_P=\frac{1}{P}\sum_{p=1}^P A_p$ (and the same expression for $B_P$), where $A_p$ and $B_p$ are the coefficients of the individual $p_{\rm{th}}$ noisy image. As $P$ increases, the training becomes more and more effective
and the regression coefficients approach their optimal values. 

\begin{figure}
\begin{center}
\includegraphics[width=7.0cm]{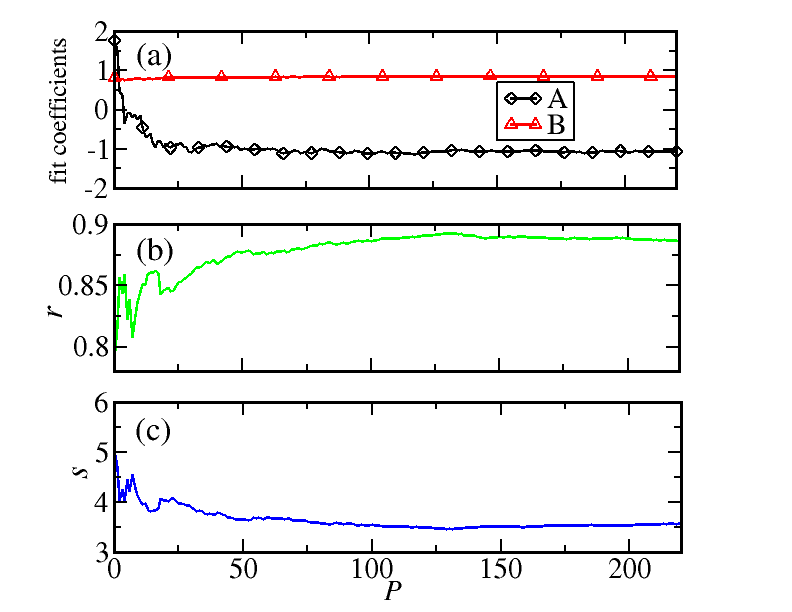}
\caption{a) Fit coefficients for the digit 0 and noise $\varepsilon=1$, b) Correlation coefficient and c) mean square error as a function of the number of samples P. The correlation coefficient remains close to one, while
the other observables quickly approach their optimal values, after about 150 iterations.}
\label{fig:figure3}
\end{center}
\end{figure}

The values of the regression coefficients for the digit 0 and noise intensity $\varepsilon=2$ as a function of the number of samples $P$ is displayed in Fig.\ref{fig:figure3}a). The quality of the fit is parametrised by the correlation coefficient $r$, displayed in Fig.\ref{fig:figure3}b), and one can see that it is very close to 1.
Fig.\ref{fig:figure3}c) shows the mean error $s=\sum_{k=1}^K\frac{[\mathcal{J}^N(t_k)-\mathcal{\overline{J}}^N]^2} {K-2}$, with $\mathcal{\overline{J}}^N$ indicating the time average of the noisy current.
One can observe here that $s$ decays exponentially and reduces by a factor two and then stabilises after about 150 iterations. 

The fundamental parameter for the evaluation of the performance of our model is its capability to distinguish between different digits. To measure this parameter, we generate an ensemble of noisy images representing various digits.
Let us take as example the digit 0 and 4, with their respective currents $\mathcal{J}^N_0$ and $\mathcal{J}^N_4$ and the target current $\mathcal{J}^T_0$ for the digit 0. For each noisy image, we evaluate its distance from the target current $\mathcal{J}_0^T$. If 
$\mathcal{L}(\mathcal{J}_0^N,\mathcal{J}_0^T)<\mathcal{L}(\mathcal{J}_4^N,\mathcal{J}_0^T)$, the recognition of the digit 0 with respect to 4 is successful. The percentage of recognised images is the recognition rate.

\begin{figure}
\begin{center}
\includegraphics[width=7.0cm]{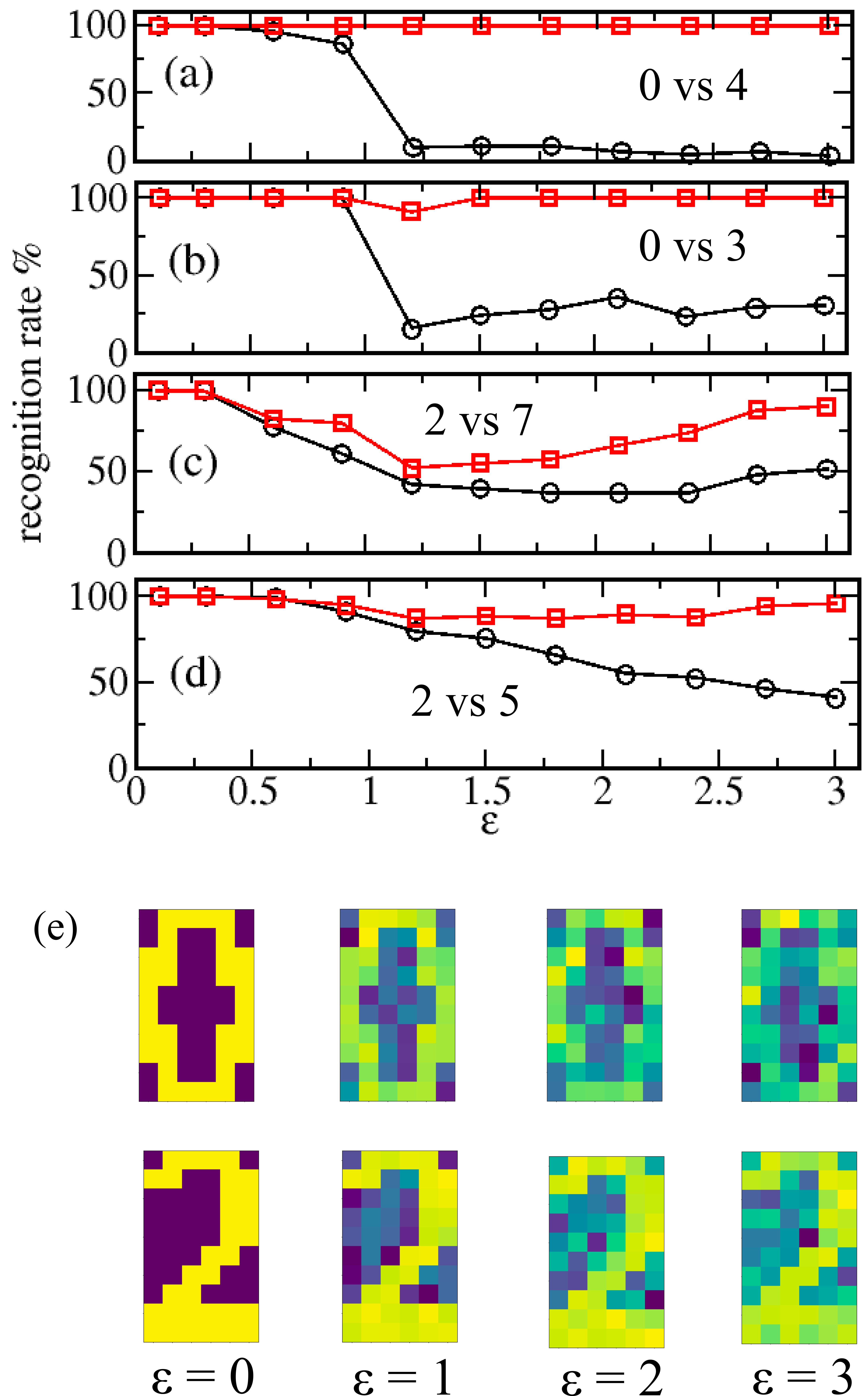}
\caption{Recognition rate for the trained (red squares) and untrained (black circles) systems as a function of the noise strength $\varepsilon$ in the chemical potential. Panels a) and b) show respectively the recognition 
rate between digit 0 and digits 4 and 3. Panels c) and d) show respectively the recognition rate between digit 2 and digits 7 and 5. Finally, panel e) displays the two digits degraded by noise.}
\label{fig:figure4}
\end{center}
\end{figure}
Fig.\ref{fig:figure4} shows the recognition rate for various digits as a function of the noise intensity, both for the trained and untrained case. We observe  that 150 samples are enough to train the system and reach a recognition rate of 100\% in the best case, up to $\varepsilon=3$, which is comparable to the performance of the human eye.

In conclusion, we have demonstrated a simple and straightforward way to perform neuromorphic computing with systems driven out of thermal equilibrium. We wish to stress that the proposed approach is very general and independent on the physical realisation,
provided that one can identify thermodynamical forces and fluxes in the system. 

So far, we have considered only a few digits, and further investigation is needed to determine whether the approach is robust and general enough to recognise more digits.
Moreover, we have considered only the zero temperature limit, where the only thermodynamical forces are differences in chemical potentials. 
However, in principle it is possible to encode the image using an arbitrary driving, including a rf field or a temperature differences. This can introduce another source of fluctuations and further analysis is 
needed to understand how those fluctuations influence the performance of the system. These issues will be discussed in a forthcoming paper.

We thank S. Iubini, S. Lepri, and G. Q. Maguire Jr. for useful discussions. 

Financial support from Vetenskapsr{\aa}det (grant numbers VR 2015-04608, VR 2016-05980 and VR 2016-01961), and Swedish Energy Agency (grant number STEM P40147-1) is acknowledged.
The computations were performed on resources provided by the Swedish National Infrastructure for Computing (SNIC) at the National Supercomputer Center (NSC), Link\"oping University, the PDC Centre for High Performance Computing (PDC-HPC), KTH, and the High Performance Computing Center North  (HPC2N), Ume{\aa} University.


\end{document}